\documentclass[aps,pra,superscriptaddress,twocolumn]{revtex4-1}
\usepackage{graphicx}
\usepackage{amsmath}
\usepackage{amssymb}
\usepackage{xcolor}
\usepackage{braket}
\usepackage{bm}
\usepackage{soul}
\usepackage{ulem}
\usepackage[colorlinks,pdfusetitle,urlcolor=blue,citecolor=blue,linkcolor=blue,bookmarksnumbered,plainpages=false]{hyperref}

\newcommand{\refA}{Mollow1969}
\newcommand{\refB}{Power1995}
\newcommand{\refC}{Milonni2015}
\newcommand{\refD}{Donaire2015c}
\newcommand{\refE}{Donaire2016a}
\newcommand{\refF}{Jentschura2017a}
\newcommand{\refG}{Jentschura2017}
\newcommand{\refH}{Barcellona2016}

\newcommand{\B}[1]{\mathbf{#1}}
\newcommand{\tens}[1]{\textbf{\textsf{#1}}}
\renewcommand{\i}{\mathrm{i}}

\newcommand{\overbar}[1]{\mkern 1.5mu\overline{\mkern-1.5mu#1\mkern-1.5mu}\mkern 1.5mu}
\newcommand{\rad}{r}

\newcommand{\res}{\text{res}}

\newcommand{\new}[1]{#1}

\begin{document}

\title{Lateral interatomic dispersion forces}

\author{Pablo Barcellona}
\email{pablo.barcellona@gmail.com}
\affiliation{Physikalisches Institut, Albert-Ludwigs-Universit\"at
Freiburg, Hermann-Herder-Str. 3, 79104 Freiburg, Germany}

\author{Robert Bennett}
\affiliation{Physikalisches Institut, Albert-Ludwigs-Universit\"at
Freiburg, Hermann-Herder-Str. 3, 79104 Freiburg, Germany}
\affiliation{School of Physics \& Astronomy, University of Glasgow, Glasgow, G12 8QQ, United Kingdom}

\author{Stefan Yoshi Buhmann}
\affiliation{Physikalisches Institut, Albert-Ludwigs-Universit\"at
Freiburg, Hermann-Herder-Str. 3, 79104 Freiburg, Germany}
\date{\today}

\begin{abstract}
Van der Waals forces between atoms and molecules are universally assumed to act along the line separating them. Inspired by recent works on effects which can propel atoms parallel to a macroscopic surface via the Casimir--Polder force, we predict a lateral van der Waals force between two atoms, one of which is in an excited state with non-zero angular momentum and the other is isotropic and in its ground state. The resulting force acts in the same way as a planetary gear, in contrast to the rack-and-pinion motion predicted in works on the lateral Casimir--Polder force in the analogous case, for which the force predicted here is the microscopic origin. We illustrate the effect by predicting the trajectories of an excited caesium in the vicinity of ground-state rubidium, finding behaviour qualitatively different to that if lateral forces are ignored. 
\end{abstract}

\maketitle

Descriptions of macroscopic phenomena are often informed and improved by understanding the underlying microscopic processes. Examples are found throughout condensed matter physics, for instance the BCS theory of superconductivity \cite{Bardeen1957} or the Lifshitz theory of Casimir forces \cite{E.M.Lifshitz1956}. The latter explains Casimir's original result \cite{Casimir1948} for the attraction between two perfectly conducting parallel plates in terms of correlations between the fluctuating charge distributions of their elementary atomic constituents. This is part of a broad class of phenomena known as dispersion interactions (c.f. \cite{Buhmann2012BothBooks}), the most familiar being the Van der Waals force between two neutral atoms. Closely related to this is the Casimir--Polder force \new{that a} neutral atom feels in proximity to a material body. 

In recent years, lateral Casimir (surface--surface) and Casimir--Polder (atom--surface) \footnote{In this work we refer to atom-atom forces at all distances as van der Waals forces (in contrast to some authors who refer to the long-distance atom-atom interaction as a Casimir-Polder force), and all atom-surface forces as Casimir-Polder forces. This should not be taken as denial of the fact that Casimir and Polder derived atom-atom \emph{and} atom surface forces at all distances in their seminal paper \cite{Casimir1948a}.} forces have received attention due to their potential to realise contactless force transmission \cite{Ashourvan2007,Nasiri2012}, as well as novel types of sensors and clocks \cite{Miri2008}. All of these works rely on corrugated surfaces \cite{Messina2009,Dalvit2008,Chen2002,Rodrigues2006,Emig2003,Dobrich2008}, gratings \cite{Lambrecht2008,Contreras-Reyes2010,Bender2014,Buhmann2016}, or gyrotopic response \cite{Polevoi1985}. A number of more recent works have discussed the intriguing possibility of engineering modes propagating along a flat, featureless planar interface \cite{Rodriguez-Fortuno2013,LeKien2016,Mueller2013,Xi2013,Lin2013,Neugebauer2014,Manjavacas2017} or nanofiber \cite{Petersen2014} in such a way that an atom or second object placed nearby will feel a force dragging it along the surface. In this Letter we will reveal the microscopic origins of this latter force.

The resonant Casimir--Polder (CP) force on an atom can be expressed in terms of the dyadic Green's tensor $\overbar{\tens{G}}\left(\mathbf{r},\mathbf{r}',\omega  \right)$ describing propagation of electromagnetic waves of frequency $\omega$ from point $\B{r}'$ to $\B{r}$ subject to boundary conditions imposed by material geometry. For a two-level atom at position $\B{r}_\text{A}$ with time-dependent excited-state occupancy $p(t)$ \new{it} is given by \cite{Scheel2015,OudeWeernink2018}
\begin{align}
\textbf{F}^\res(\textbf{r}_\text{A},t) &=   2\mu _0 p(t) \omega_\text{A}^2 \notag \\ &\quad \times \text{Re}\Big[ \nabla  \textbf{d}_{10}^\text{A} \cdot \overbar{\tens{G}}\left( \textbf{r},\textbf{r}_\text{A},\omega_\text{A} \right) \cdot \textbf{d}_{01}^\text{A}  \Big]_{\textbf{r} = \textbf{r}_\text{A}},
\label{CasPol}
\end{align}
where $\omega_\text{A}$ is the transition frequency and $ \textbf{d}_{01}^\text{A}=\textbf{d}_{10}^{\text{A}*} $ is the (complex) transition dipole moment from the upper to lower level, and $\mu_0$ is the permeability of free space. \new{There is also a non-resonant force originating in the contribution from photons with frequencies different to the atomic transition, but as shown in the Supplementary Material \footnote{See Supplementary Material [url] for detailed derivations of the forces, emission rates and emission spectra.} the contribution of this for the parameters we will choose is negligible compared to the resonant terms.}  Most derivations of Casimir--Polder forces proceed by finding the position-dependent energy shift of the atomic levels, then taking a spatial derivative to find the force. If the atom has a complex polarisability (and corresponding complex dipole moment) then the Casimir--Polder force is not conservative, meaning that it cannot be derived as the gradient of an energy shift. We seek a microscopic version of the non-conservative force given by Eq.~\eqref{CasPol}, which was derived from the Lorentz force law.

From a microscopic point of view, a macroscopic medium is a collection of a large number of atoms --- the imposition of macroscopic boundary conditions is simply a neat and powerful way of summarising their collective behaviour. We thus begin by replacing the material body found in accounts of the lateral Casimir--Polder force with a collection of neutral atoms. This is done by taking the dilute-gas limit (\new{in which the polarisability volume of each atom is much smaller than the cube of the mean interatomic spacing}) in a similar manner to that done by Lifshitz \cite{E.M.Lifshitz1956} via a Born-expansion of the dyadic Green's tensor (see, for example, \cite{Purcell1973,Buhmann2006,Sherkunov2007})
\begin{align}\label{BornExpansion}
&\overbar{\tens{G}}\left(\mathbf{r},\mathbf{r}',\omega  \right) = {\tens{G}}\left( \mathbf{r},\mathbf{r}',\omega  \right)\notag  \\
&\!\!+ \mu _0 \omega ^2\!\!\int \! \text{d}^3 r'' \!\rho \left( \mathbf{r}'' \right){\tens{G}}\left( \mathbf{r},\mathbf{r}'',\omega  \right) \cdot \bm{\alpha}_\text{B} \left( \omega  \right)  \cdot {\tens{G}}\left( \mathbf{r}'',\mathbf{r}',\omega  \right) \new{+ \ldots} \end{align}
where $\rho(\B{r})$ is the number density of a collection of arbitrarily-placed atoms with identical polarisibilities $\bm{\alpha}_\text{B}\left( \omega  \right)$, and ${\tens{G}}\left( \mathbf{r},\mathbf{r}',\omega  \right)$ is the known Green's tensor of the background environment which could for example be unbounded vacuum, but need not be. 

Using the Born-expanded Green's tensor \eqref{BornExpansion} \new{with a delta-distributed number density} in the expression \eqref{CasPol} for the \new{resonant} force, one finds that $\textbf{F}^\res(\textbf{r}_\text{A},t)=\bar{\textbf{F}}^\res(\textbf{r}_\text{A},t) + \int d^3 r' \rho (\B{r}') \B{F}^\res(\B{r}_\text{A},\B{r}',t)$, where $\bar{\textbf{F}}^\res(\textbf{r}_\text{A},t)$ is the force felt between atom A and the background bodies alone, and;
\begin{align}\label{ForceA}
\textbf{F}^\res(\textbf{r}_\text{A},\B{r}',t)=  & 2\mu _0^2p(t)\omega_\text{A}^4 \text{Re}  \big[\nabla\textbf{d}_{10}^\text{A} \cdot {\tens{G}}\left( \textbf{r},\textbf{r}',\omega_\text{A} \right)\notag \\
&   \cdot \bm{\alpha}_\text{B}(\omega_\text{A})\cdot {\tens{G}}\left( \textbf{r}',\textbf{r}_\text{A},\omega_\text{A} \right) \cdot \textbf{d}_{01}^\text{A} \big] _{\textbf{r} = \textbf{r}_\text{A}}.
\end{align}
This is an atom-atom (van der Waals) force felt by atom A due to the presence of a (non-identical) atom B at $\B{r}'=\B{r}_\text{B}$ with dynamic polarisability tensor $\bm{\alpha}_\text{B}(\omega)$, valid as long the atoms are far enough apart that there is no appreciable wave-function overlap. Equation \eqref{ForceA} is made up of both the interaction of atom A with its own field as reflected by atom B, and the interaction with the quantised electromagnetic vacuum field. For most naturally-arising situations, the atomic dipoles can be considered to be randomly oriented, leaving an average force which pulls the particles linearly together (or, in some rare cases, pushes them apart).

\begin{figure}
\includegraphics[width=\columnwidth]{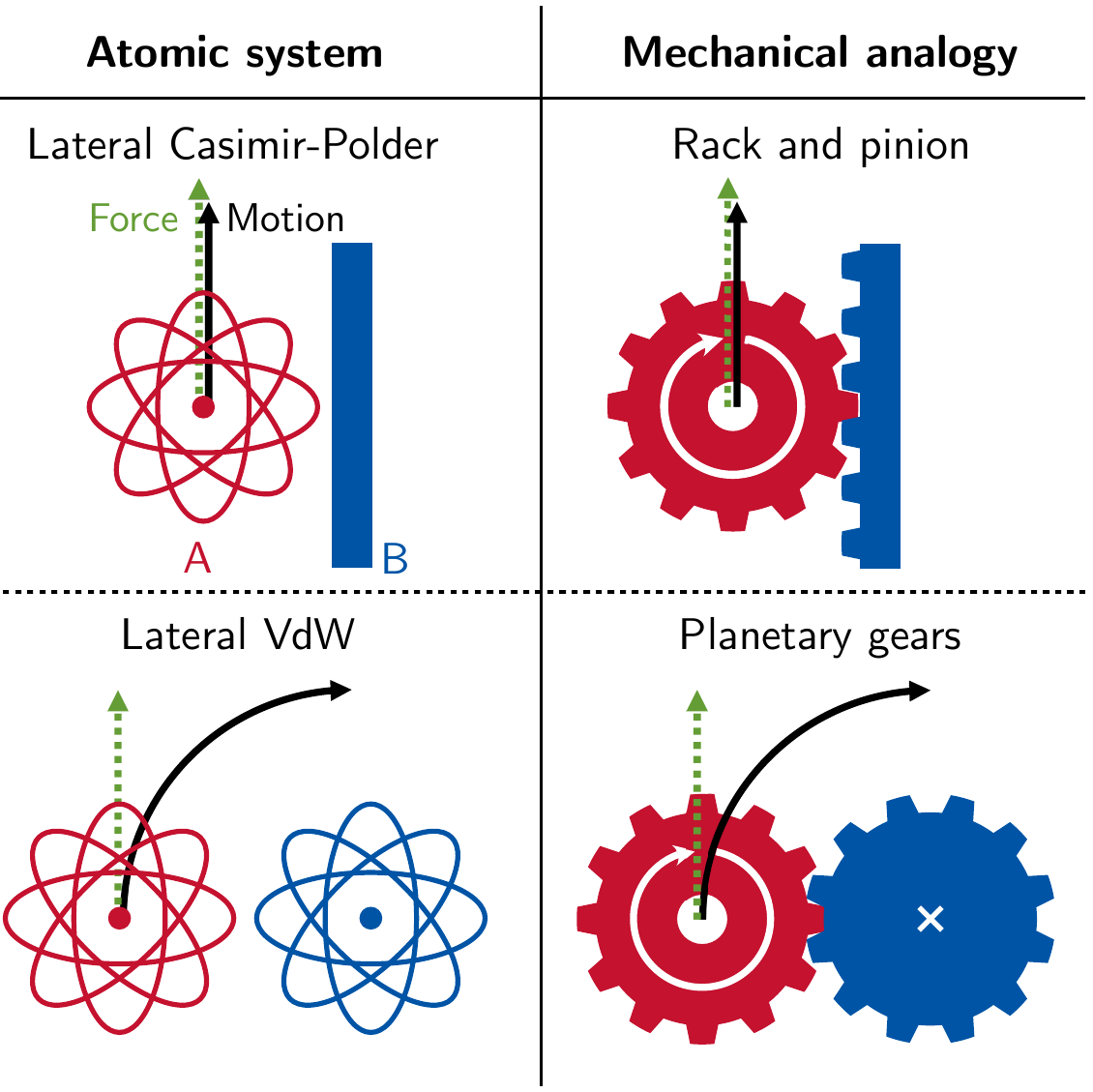}
\caption{Mechanical analogies to the lateral Casimir--Polder force studied in previous works, and the lateral interatomic force discussed here. Dashed (green) arrows represent forces, while solid arrows (black, white) represent motion. In all cases the entity on the right (blue) is considered as being fixed in space. }
\label{Mechanical}
\end{figure}

The situation changes drastically if one of the atoms has a complex dipole moment, corresponding to an atomic transition with different magnetic quantum numbers --- loosely thought of as a continuous rotation. As we will show, the resulting force causes atom A to orbit atom B. Extending the analogy of the lateral Casimir--Polder force with a rack and pinion to our situation, the interaction considered here could be considered as an atom-scale, contactless version of planetary gearing as illustrated in Fig.~\ref{Mechanical}.

We will illustrate this by taking atom A to be caesium undergoing a D2 transition from the highest hyperfine state $\ket{6^2\text{P}_{3/2}, F=5, M_\text{F} = 5} \equiv \ket{1}$ to the hyperfine ground state $\ket{6^2\text{S}_{1/2}, F=4, M_\text{F} = 4}\equiv \ket{0}$, and atom B to be rubidium in its ground state (\new{$5^2\text{S}_{1/2}$}, polarizability $\bm{\alpha}_\text{B} = \alpha_\text{B} \text{diag}(1,1,1)$, where $\alpha_\text{B} =   4\pi \varepsilon_0 \times 293\mathrm{\AA}^3$ at the caesium D2 wavelength of 852nm \cite{Sansonetti2005,SteckData}). The magnitude of the transition dipole moment between these two caesium levels is ${d}_\text{A} \equiv |\B{d}^\text{A}_{10}| = 2.68 \times 10^{-29}$Cm \cite{Scheel2015,SteckData}, while its components in the lab frame depend on the character of the light which excites the transition. Assuming a right-circularly polarised laser beam propagates along the $y$ direction of a cartesian co-ordinate system, the transition dipole moment can be written as:
\begin{equation}\label{DipoleMoment}
\B{d}^\text{A}_{10} = \frac{d_\text{A}}{\sqrt{2}} (\i,0,1) \; . 
\end{equation}
We assume that \new{the atoms are in free space, with} atom B at the origin and atom A in the $xz$ plane at position $z = \rad_\text{A} \cos\theta_\text{A}$, $x = \rad_\text{A} \sin \theta_\text{A}$. The two lateral components of the  \new{resonant} force are in the $\theta$ and $y$ directions, and are found by inserting the free-space Green's tensor $\tens{G}^{(0)}$ into \eqref{ForceA}. As shown in, e.g., Ref.~\cite{Buhmann2012BothBooks}, this is given explicitly by
\begin{equation}
\tens{G}^{(0)}\left( \mathbf{r},\mathbf{r}',\omega  \right) = \left( \tens{I} + \frac{c^2}{\omega ^2}\nabla \nabla  \right)\frac{e^{\mathrm{i}\omega \left| \mathbf{r} - \mathbf{r}' \right|/c}}{4\pi \left| \mathbf{r} - \mathbf{r}' \right|}
\label{G0Eq}
\end{equation}
where $c$ is the speed of light. Using cylindrical coordinates $\mathbf{r}=\left(r\sin \theta,y,r\cos \theta \right)$, $\nabla f= \frac{\partial f}{\partial r} \hat{\bm{r}} + \frac{1}{r} \frac{\partial f}{\partial \theta} \hat{\bm{\theta}} + \frac{\partial f}{\partial y} \hat{\bm{y}}  $ we find that the $y$ component of the force \new{$F^\res_y = \B{F}^\res\cdot \hat{\B{y}}$} vanishes;
\begin{equation}\label{lateralfy}
 F^\res_y(r_\text{A},t)=0
\end{equation}
and the $\theta$ component \new{$F^\res_\theta = \B{F}^\res\cdot \hat{\bm{\theta}}$} is:
\begin{equation}F^\res_\theta(\rad_\text{A},t) =  - \frac{p(t)}{40\pi ^2\varepsilon _0^2c^5\rad_\text{A}^2}d_{\text{A}}^2\alpha_\text{B}(\omega _{\text{A}})\omega _{\text{A}}^5 g \left( \frac{\omega_{\text{A}} \rad_\text{A}}{c} \right)
 \label{lateralf}
\end{equation}
where $\varepsilon_0$ is the permittivity of free space and we have defined
\begin{align}
g(\eta) \equiv\frac{5}{2\eta^5}[ 6\eta\left( \eta^2 - 3 \right) &\cos (2\eta) \notag \\ &+ \left( 9 - 15\eta^2 + \eta^4 \right)\sin (2\eta) ].
\end{align}
\new{The lateral force shown in Eq.~\eqref{lateralf} is our main result, but as a point of comparison we also report the normal force} $F^\res_r = \B{F}^\res\cdot \hat{\bm{r}}$:
\begin{align}
F^\res_r (r_{\text{A}},t) =&-  \frac{15 p \left(t \right)}{16\pi ^2\varepsilon _0^2 r_\text{A}^7 }d_{\text{A}}^2 \alpha_{\text{B}}(\omega _{\text{A}}) h \left( \frac{\omega_{\text{A}} r_{\text{A}}}{c} \right), \label{Fr}
\end{align}
where:
\begin{align}\label{hOfEta}
h \left(\eta \right) =&\frac{1}{15}\Big[ 3\left(5-8 \eta^2 + \eta^4 \right)\cos (2\eta)\notag \\
& +\eta  \left( 30 - 10\eta^2 + \eta^4 \right)\sin (2\eta) \Big].
\end{align}
\new{Similar normal forces between (non-rotating) excited and ground state atoms are well-studied, having been considered by the authors of Refs~\cite{\refB,\refC,\refD,\refE,\refF,\refG,\refH} with particular emphasis on the oscillating distance dependence, but the lateral force \eqref{lateralf} predicted here has not previously been discussed.} The van der Waals interaction in the near field \new{(non-retarded)} limit $\omega_\text{A} r_\text{A}/c \ll 1$, is given by  Eqs.~(\ref{lateralf})  and (\ref{Fr}),  where $\lim _{\eta  \to 0}g\left( \eta  \right) = 1, \lim _{\eta  \to 0}h\left( \eta  \right) = 1$, \new{while the far-field (retarded) limit is found from Eqs.~(\ref{lateralf})  and (\ref{Fr}) by taking  $\omega_\text{A} r_\text{A}/c \gg 1$}. It is interesting to note that the forces are independent of $\theta_\text{A}$ which also results from symmetry considerations. Formulae (\ref{lateralf}) and (\ref{Fr}) account for retardation effects via the function $g$ in the limit $\omega_{\text{A}} \rad_\text{A}/c  \gg 1$, which arises because of the finite velocity of light. In the retarded regime the time taken for the photon to reach the second atom and reflect back to the first atom become comparable with the time scale of the dipole fluctuations themselves. In this case the orientation of the dipole at the time of emission may differ from its orientation at the time of absorption of the reflected photon, reducing the attractive force as compared to the ideal case of parallel alignment.

Our next step is to recognise that the excited-state interatomic force can be understood as a recoil force originating from the exchange of excitations with the environment, \new{for which we present an alternative derivation of Eq.~\eqref{ForceA} [and thereby Eqs \eqref{lateralfy} and \eqref{lateralf}], based on emission spectra instead of forces \cite{Sherkunov2009}.}
\new{To do this we begin by calculating the spontaneous decay rate for atom A in the excited  state $\left| 1\right\rangle $ in the presence of a second atom B. As shown explicitly in the supplementary material, in free space this is given by;}
\begin{align}\label{GammaRShifted}\Gamma(\textbf{r}_\text{A},\textbf{r}_\text{B})=  \frac{ 2\mu _0^2}{\hbar}  \omega _\text{A}^4 \text{Im} \Big[  \textbf{d}_{10}^\text{A} \cdot \tens{G}^{(0)}\left( \textbf{r}_\text{A},\textbf{r}_\text{B},\omega _\text{A} \right)\notag \\
\cdot \bm{\alpha}_\text{B} (\omega_\text{A})\cdot \tens{G}^{(0)}\left( \textbf{r}_\text{B},\textbf{r}_\text{A},\omega _\text{A} \right) \cdot \textbf{d}_{01}^\text{A} \Big] \, .
\end{align}

\new{We can define a momentum-space emission rate density $\gamma$ as
\begin{equation}
\Gamma(\mathbf{r}_\text{A},\mathbf{r}_\text{B})= \int \text{d}^3k \gamma \left(\mathbf{k}; \mathbf{r}_\text{A}, \mathbf{r}_\text{B}\right),
\end{equation}
which is  the rate at which light with wavevector $\mathbf{k}$ is emitted, if the atom $\text{A}$ is in the excited state.
Since the free-space Green's tensor can be Fourier transformed $\tens{G}^{(0)} \left( \textbf{r},\textbf{r}',\omega\right)=\left(2\pi\right)^{-3}\int \text{d}^3k \text{e}^{\text{i} \mathbf{k} \cdot \left(\mathbf{r}-\mathbf{r}'\right)} \tens{G}^{(0)} \left( \mathbf{k},\omega\right)$ the rate density reads:
\begin{align}\label{GammaRDef}
\gamma(\mathbf{k}; \textbf{r}_\text{A},\textbf{r}_\text{B})= & \frac{ 2\mu _0^2}{\left(2\pi\right)^3\hbar}  \omega _\text{A}^4 \text{Im} \Big[\text{e}^{\text{i} \mathbf{k} \cdot \left(\mathbf{r}_\text{A}-\mathbf{r}_\text{B}\right)}   \textbf{d}_{10}^\text{A} \cdot \tens{G}^{(0)}\left( \mathbf{k},\omega _\text{A} \right) \notag  \\
&\cdot {\bm{\alpha}}_\text{B} (\omega_\text{A})\cdot  \tens{G}^{(0)}\left( \textbf{r}_\text{B},\textbf{r}_\text{A},\omega _\text{A} \right) \cdot \textbf{d}_{01}^\text{A} \Big].
\end{align}}
\new{Explicit evaluation of the rate density in our particular setup (see supplemental material) reveals that $\gamma(-\mathbf{k}; \textbf{r}_\text{A},\textbf{r}_\text{B}) \ne \gamma(\mathbf{k}; \textbf{r}_\text{A},\textbf{r}_\text{B})$, showing that the net recoil force is, as expected, not zero.} \new{This can be explained by noting that the momentum-space recoil force density is given by $-\gamma \hbar  \mathbf{k}$  (the minus signs accounting for the fact that we are considering recoils), so that the total resonant force on atom A is given by
\begin{equation}\label{FResFromGamma}
\mathbf{F}^{\text{res}} \left(\mathbf{r}_\text{A},\mathbf{r}_\text{B},t\right)=-p\left(t\right)  \int \text{d}^3k \hbar \mathbf{k} \gamma \left(\mathbf{k}; \mathbf{r}_\text{A}, \mathbf{r}_\text{B}\right).
\end{equation}
Since $\nabla \text{e}^{\text{i} \mathbf{k}\cdot \mathbf{r}}=\text{i} \mathbf{k} \text{e}^{\text{i} \mathbf{k}\cdot \mathbf{r}} $ we immediately find the recoil force Eq.~\eqref{ForceA}, which leads to the lateral forces \eqref{lateralfy} and \eqref{lateralf}. }

\begin{figure}
\includegraphics[width=\columnwidth]{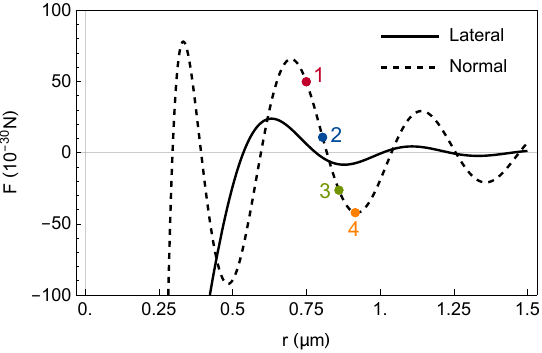}
\caption{Lateral [solid, Eq.~\eqref{lateralf}] and normal [dashed, Eq.~\eqref{Fr}] \new{resonant} forces on a caesium atom (D2 transition) due to the presence of a rubidium atom at the origin. The numbered dots are those used later for trajectory simulations. Each chosen distance is comfortably larger than the atomic radii ($\sim10$\AA), consistent with our assumption of independent polarisibilities. }
\label{Plot1D}
\end{figure}

We are now left with a remarkable conclusion. The asymmetry that atom B represents in the environment of atom A causes the latter to preferentially release its excitation in a direction perpendicular to the line joining them, propelling A around B like a planetary gear. When combined with the oscillatory nature of the  \new{resonant} force that atom B exerts on atom A, we also find that the sign of this torque can be varied by changing the distance between the atoms, as shown in Fig.~\ref{Plot1D}, 
where we also plot the corresponding normal  \new{resonant} force \eqref{Fr}.

\begin{figure}
\centering
\includegraphics[width=0.8\columnwidth]{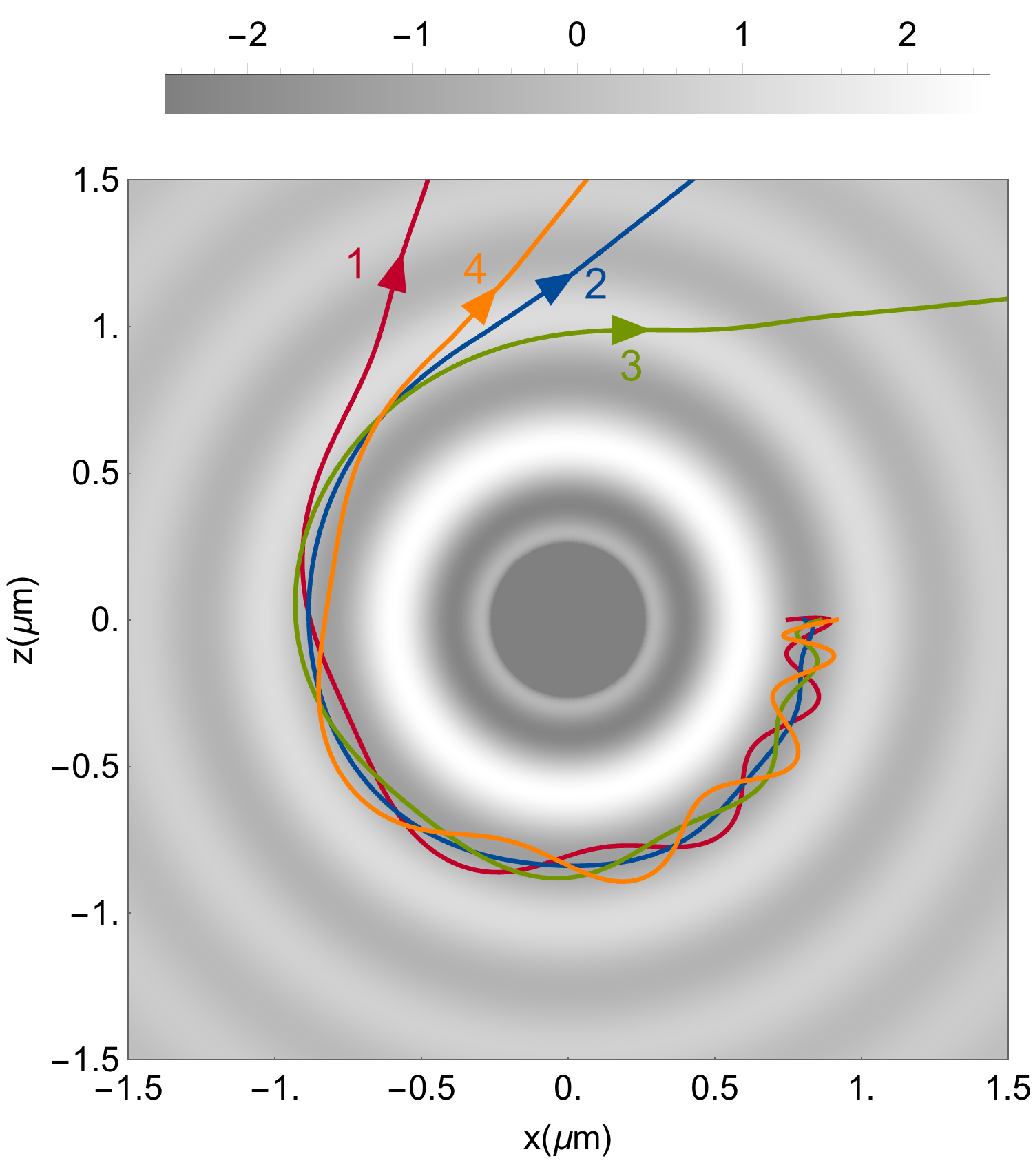}
\caption{Simulated trajectories for a caesium atoms starting at rest for the four points shown in Fig.~\ref{Plot1D}. Shown in the background is the potential energy function found by integrating the normal  \new{resonant} force in the radial direction.}
\label{StreamPlot}
\end{figure}

\begin{figure}
\includegraphics[width=\columnwidth]{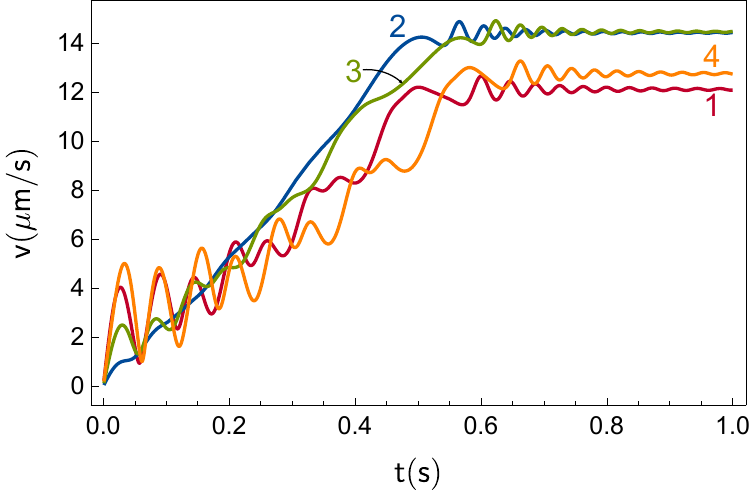}
\caption{Velocities gained along the four trajectories simulated in Fig.~\ref{StreamPlot}. }
\label{VelocityPlot}
\end{figure}

Having seen that a lateral interatomic dispersion force is possible, we now turn our attention to its magnitude and prospects for experimental observation. In the absence of external driving, the atomic population (and therefore the recoil force) decays on average like $e^{-\Gamma t}$, meaning that the torque quickly becomes unobservably small. In order to combat this, we \new{introduce a coherent driving, for which it us useful to go} into the vacuum picture where the interaction of an atom with a coherent field can be considered as being made up of a classical driving field plus the vacuum field \cite{Pegg1980,Dutra1994,Fuchs2018c}. We consider atom A to be continuously driven by a circularly-polarised classical laser field propagating in the positive $y$-direction: 
\begin{equation}\label{DrivingLaser}
\mathbf{E}_{\text{L}}\left( t \right) = E_0\B{e}_{\text{R}} e^{ - \text{i} \omega _{\text{L}}t}/2 + \text{c.c.}
\end{equation}
where $E_0$ is the field's amplitude, $\omega_{\text{L}}$ its frequency and $\B{e}_{\text{R}}= \left(-\text{i},\; 0, \; 1 \right)/\sqrt 2$. The effect of the driving laser is accounted for by the real Rabi frequency ${ \hbar \Omega  = \mathbf{d}_{10}^{\text{A}} \cdot \B{e}_{\text{R}} E_0=  d_\text{A} E_0 }$. Solving the optical Bloch equations \new{for the interaction of the laser field with atom A in the absence of atom B} in the long time limit ($t\gg \Gamma^{-1}$), the expectation value of the dipole moment operator of atom A is then given by
\begin{align}\label{DipoleExpct}
\left\langle \mathbf{d}^\text{A}(t) \right\rangle  &=   \frac{\sqrt{2}\Omega \Delta }{2\Delta ^2 + \Omega^2} d_\text{A}\big( \sin \left( \omega _\text{L}t \right),0, -\cos \left( \omega _\text{L}t \right) \big)
\end{align}
where $\Delta = \omega_\text{L}-\omega_\text{A}$ is the detuning of the laser field from the atomic resonance, and we have also assumed $\Gamma \ll |\Delta|$. In the absence of atom B, atom A simply rotates in the $x-z$ plane with the same frequency as the laser, which is not surprising. The presence of atom B breaks the symmetry of the electromagnetic environment experienced by atom A. To quantify this effect we use Eq.~\eqref{ForceA} with an excited state population given by (\new{see, for example, \cite{\refA}});
\begin{equation}\label{Prob}
p(t) =\frac{ \Omega^2}{4\Delta ^2 + 2\Omega^2}
\end{equation}
In the strong interaction limit $\Omega \gg |\Delta|$, the effect of the resultant force in is shown in Fig.~\ref{StreamPlot}, 
where we place atoms initially at rest on the $x$ axis at the positions indicated by the dots in Fig.~\ref{Plot1D} and compute their trajectories. \new{The illuminating light should be set up in such a way that it has a constant amplitude over the trajectory of atom A, while affecting atom B as little as possible. This could be achieved, for example, by tailoring atom B's level structure, or through the use of structured light.}  It is seen that under such continuous laser driving the lateral force causes atom A to be ejected after slightly more than half an orbit of the fixed, isotropic atom B. In Fig.~\ref{VelocityPlot}
we plot the velocity gained as a function of time, finding $12-15\mu$m/s for the parameters chosen here. To reach these velocities takes a relatively long time (on the order of a second) since the force is so weak. However, there are several routes to combat this by enhancement of the interaction. One might expect that use of Rydberg atoms with their large dipole moments (quadratic in the principle quantum number $n$), however the energy difference of adjacent states scaled as $n^{-3}$ meaning that the force derived here is strongly suppressed for such systems. Finally, we note that the interaction could be enhanced by placing the pair of atoms in a cavity, in much the same was as the spontaneous decay rate of a quantum emitter can be enhanced through the Purcell factor \cite{Purcell1946}. 

To conclude, we have demonstrated the existence of a lateral Van der Waals force on an excited, circularly polarised atom due to the placement of an isotropic, ground state atom nearby. We have outlined how the effect might be experimentally accessed by selectively pumping the atom to a Zeeman sub-level. Control of the lateral force direction and magnitude can be experimentally implemented by changing the handedness of the illuminating light and the distance between the two atoms. Our work is the first demonstration of the most elementary lateral force that can act on a circularly polarised emitter, without the influence of a surface. Nevertheless, our expression of the force in terms of the dyadic Green's tensor means that additional macroscopic objects can be introduced without fundamental changes to the method, opening up the effect detailed here to Purcell-type enhancement. In the longer term, the force could find applications in optomechanics as a new actuation method, as well as in any of the numerous fields in which Van der Waals forces play a pivotal role.

\acknowledgments{\new{The authors thank Gabriel Dufour for valuable feedback on the manuscript, and the Deutsche Forschungsgemeinschaft for financial support (grant BU 1803/3-1476)}.}

\end{document}